\documentclass[12pt,a4paper,american]{article}
\usepackage{amsmath}
\usepackage{graphicx}
\usepackage{amssymb}
\usepackage{esint}

\makeatletter
\newcommand{\lyxaddress}[1]{
\par {\raggedright #1
\vspace{1.4em}
\noindent\par}
}

\usepackage{amsthm}

\usepackage{mathrsfs}

\theoremstyle{remark}

\newtheorem*{remark*}{Remark}
\newcommand{\sF}{\mathscr{F}}
\newcommand{\sH}{\mathscr{H}}
\newcommand{\sK}{\mathscr{K}}

\newcommand{\diag}{\mathop\mathrm{diag}\nolimits}
\newcommand{\offdiag}{\mathop\mathrm{offdiag}\nolimits}
\newcommand{\Ran}{\mathop\mathrm{Ran}\nolimits}

\newcommand{\sign}{\mathop\mathrm{sign}\nolimits}

\makeatother

\usepackage{babel}

\begin{document}

\title{A charged particle in a homogeneous magnetic field accelerated by
a time periodic Aharonov-Bohm flux}

\date{{}}

\author{T.~Kalvoda$^{1}$, P.~\v{S}\v{t}ov\'\i\v{c}ek$^{2}$}

\maketitle

\lyxaddress{$^{1}$Department of Theoretical Computer Science, Faculty of Information
Technology, Czech Technical University in~Prague, Kolejn\'\i~2,
160~00 Praha, Czech Republic}

\lyxaddress{$^{2}$Department of Mathematics, Faculty of Nuclear Science, Czech
Technical University in~Prague, Trojanova 13, 120~00 Praha, Czech
Republic}
\begin{abstract}
\noindent We consider a nonrelativistic quantum charged particle moving
on a plane under the influence of a uniform magnetic field and driven
by a periodically time-dependent Aharonov-Bohm flux. We observe an
acceleration effect in the case when the Aharonov-Bohm flux depends
on time as a sinusoidal function whose frequency is in resonance with
the cyclotron frequency. In particular, the energy of the particle
increases linearly for large times. An explicit formula for the acceleration
rate is derived with the aid of the quantum averaging method, and
then it is checked against a numerical solution with a very good agreement.\\

\end{abstract}

\section{Introduction}

The problem of acceleration in physical systems driven by time-periodic
external forces, both in classical and quantum mechanics, has a rather
long history though the results in the latter case are much less complete.
One of the most prominent examples which initiated a lot of efforts
in this field is the so called Fermi accelerator. On the basis of
a theory due to Fermi to explain the acceleration of cosmic rays \cite{Fermi}
Ulam formulated a mathematical model describing a massive particle
bouncing between two infinitely heavy walls while one of the walls
is oscillating \cite{Ulam}. A thorough analysis finally did not fully
confirm the expectations, however \cite{ZaslavskiiChirikov,LiebermanLichtenberg,Pustylnikov}.
The model has also been reformulated in the framework of quantum mechanics
\cite{Karner}.

More models of this sort have been studied in detail so far but we
just mention one of them, the so called electron cyclotron resonance.
One readily finds that electrons in a uniform magnetic field can gain
energy from a microwave electric field whose frequency is equal to
the electron cyclotron frequency. Because of an unlimited energy increase
the relativistic effects cannot be neglected in a complete analysis.
But even the relativistic model admits a quite explicit characterization
of the resonant solution for a transverse circularly polarized electromagnetic
wave propagating along the uniform magnetic field \cite{RobertsBuchsbaum}.
In experimental arrangements the heated electrons are confined in
a magnetic mirror field. Consequently, as they move along a flux tube
of the mirror field they are exposed to the resonance heating only
in a restricted region \cite{Seidl,Grawe,JaegerLichtenbergLieberman}.
This acceleration mechanism is widely used in plasma physics.

Here we wish to discuss, on the quantum level, a model sharing some
features with the preceding one. We again consider a charged particle
placed in a uniform magnetic field. In our model the situation is
simplified, however, in the sense that the particle is confined to
a plane perpendicular to the magnetic field. Instead of a transverse
electromagnetic wave propagating along the uniform field we apply,
as an external force, an oscillating Aharonov-Bohm flux. The frequency
of oscillations $\Omega$ again coincides with the cyclotron frequency
$\omega_{c}$ or, more generally, it may be an integer multiple of
$\omega_{c}$.

The Aharonov-Bohm effect itself received a tremendous attention as
a genuinely quantum phenomenon \cite{AharonovBohm}, and almost all
its possible aspects have been studied in the time-independent case.
For example, a careful analysis can be found in \cite{Ruijsenaars}.
On the other hand, the time-dependent case represents an essentially
more difficult mathematical problem and it has been treated so far
only marginally in a few papers \cite{Lee_etal,AgeevDavydovChirkov,AschHradeckyStovicek,AschStovicek}.

The model we propose has already been studied in the framework of
classical mechanics \cite{AschKalvodaStovicek}. It turns out that
a resonance acceleration again exists but it has some remarkable new
features if compared to the standard electron cyclotron resonance.
If $\Omega$ is an integer multiple of $\omega_{c}$, then the classical
trajectory eventually reaches an asymptotic domain where it resembles
a spiral whose circles pass very closely to the singular flux line
and, at the same time, their radii expand with the rate $t^{1/2}$
as $t$ approaches infinity. The particle moves along the circles
approximately with frequency $\omega_{c}$ while its energy increases
linearly with time. Denoting by $\mathcal{E}(t)$ the energy depending
on time, an important characteristic of the dynamics is the acceleration
rate which is computed in \cite{AschKalvodaStovicek} and is given
by the formula\begin{equation}
\gamma_{\text{acc}}:=\lim_{t\to\infty}\frac{\mathcal{E}(t)}{t}=\frac{e\omega_{c}}{4\pi}\,|\Phi'(\tau)|.\label{eq:acc_rate_class}\end{equation}
Here $\tau$ is a real number which is expressible in terms of some
asymptotic parameters of the trajectory.

The purpose of the current paper is to demonstrate that one can derive
a formula analogous to (\ref{eq:acc_rate_class}) also in the framework
of quantum mechanics. To this end and because of complexity of the
problem, we restrict ourselves to the case when the AB flux depends
on time as a sinusoidal function. To this system we apply the quantum
averaging method getting this way an approximate time evolution for
which we observe a resonance effect whose principal characterization
is again a linear increase of energy.

Let us now be more specific. We consider a quantum point particle
of mass $M$ and charge $e$ moving on the plane in the presence of
a homogeneous magnetic field of magnitude $B$. For definiteness,
all constants $M$, $e$, $B$ are supposed to be positive. Assume
further that the particle is driven by an Aharonov-Bohm magnetic flux
concentrated along a line intersecting the plane in the origin and
whose strength $\Phi(t)$ is oscillating with frequency $\Omega$.

In the time-independent case, the Hamiltonian corresponding to a homogeneous
magnetic field and a constant Aharonov-Bohm flux of magnitude $\Phi_{0}$
reads\[
\frac{\hbar^{2}}{2M}\!\left(-\frac{1}{r}\,\partial_{r}r\partial_{r}+\frac{1}{r^{2}}\!\left(-i\partial_{\theta}-\frac{e\Phi_{0}}{2\pi\hbar}+\frac{eBr^{2}}{2\hbar}\right)^{\!2}\right)\]
where $(r,\theta)$ are polar coordinates on the plane, and the Hilbert
space in question is $L^{2}(\mathbb{R}_{+}\times S^{1},r\mbox{d}r\mbox{d}\theta)$.
Making use of the rotational symmetry of the model we restrict ourselves
to a fixed eigenspace of the angular momentum $J_{3}=-i\hbar\partial_{\theta}$
with an eigenvalue $j_{3}\hbar$, $j_{3}\in\mathbb{Z}$. Put \[
p:=j_{3}-e\Phi_{0}/(2\pi\hbar).\]
Then this restriction leads to the radial Hamiltonian

\begin{equation}
H(p)=\frac{\hbar^{2}}{2M}\!\left(-\frac{1}{r}\,\partial_{r}r\partial_{r}+\frac{1}{r^{2}}\!\left(p+\frac{eBr^{2}}{2\hbar}\right)^{\!2}\right)\label{eq:def_H_p}\end{equation}
in $\sH=L^{2}(\mathbb{R}_{+},r\mbox{d}r)$. Without loss of generality,
we can assume that $p>0$ (note that $H(-p)-H(p)$ is a constant).
The boundary conditions at the origin are chosen to be the regular
ones (then $H(p)$ is the so called Friedrichs self-adjoint extension
of the symmetric operator defined on compactly supported smooth functions).
Let us note that if $0<p<1$, then more general boundary conditions
are admissible \cite{ExnerStovicekVytras} but here we confine ourselves
to the above standard choice.

Let \[
\omega_{c}=eB/M\]
be the cyclotron frequency. The operator $H(p)$ has a simple discrete
spectrum, the eigenvalues are\begin{equation}
E_{n}(p)=\hbar\omega_{c}(n+p+1/2),\quad n=0,1,2,\ldots,\label{eq:E_n_p}\end{equation}
with the corresponding normalized eigenfunctions \begin{equation}
\varphi_{n}(p;r)=c_{n}(p)\, r^{p}\, L_{n}^{(p)}\!\left(\frac{eBr^{2}}{2\hbar}\right)\exp\!\left(-\frac{eBr^{2}}{4\hbar}\right)\label{eq:basis_varphi_n}\end{equation}
where \[
c_{n}(p)=\left(\frac{eB}{2\hbar}\right)^{\!(p+1)/2}\left(\frac{2\, n!}{\Gamma(n+p+1)}\right)^{\!1/2}\]
are the normalization constants and $L_{n}^{(p)}$ are the generalized
Laguerre polynomials.

Thus our main goal is to study the time evolution governed by the
periodically time-dependent Hamiltonian $H(a(t))$ where \[
a(t)=p+\epsilon f(\Omega t)\]
and $f(t)$ is a $2\pi$-periodic continuously differentiable function,
$\Omega>0$ is a frequency and $\epsilon$ is a small parameter. This
means that the Aharonov-Bohm flux is supposed to depend on time as\begin{equation}
\Phi(t)=\Phi_{0}-(2\pi\hbar\epsilon/e)\, f(\Omega t).\label{eq:Phi_t}\end{equation}
Without loss of generality one can assume that \begin{equation}
\int_{0}^{2\pi}f(t)\mbox{d}t=0.\label{eq:aver_f_0}\end{equation}
As discussed in \cite{AschHradeckyStovicek}, for the values $0<p<1$
the domain of $H(a(t))$ in fact depends on $t$, and this feature
makes the discussion from the mathematical point of view a bit more
complicated. Nevertheless, the time evolution is still guaranteed
to exist.

\section{The Floquet operator and the quasienergy\label{sec:Floquet_quasienergy}}

Let $U(t,t_{0})$ be the propagator (evolution operator) associated
with $H(a(t))$; it is known to exist \cite{AschHradeckyStovicek}.
An important characteristic of the dynamical properties of the system
is the time evolution over a period which is described by the Floquet
(monodromy) operator $U(T,0)$, with $T=2\pi/\Omega$. We are primarily
interested in the asymptotic behavior of the mean value of energy
\[
\langle U(T,0)^{N}\psi,H(p)U(T,0)^{N}\psi\rangle\]
for an initial condition $\psi$ as $N$ tends to infinity while focusing
on the resonant case when \begin{equation}
\Omega=\mu\omega_{c}\ \,\text{for some}\ \mu\in\mathbb{N}.\label{eq:resonance}\end{equation}

A basic tool in the study of time-dependent quantum systems is the
quasienergy operator \[
K=-i\hbar\partial_{t}+H(a(t))\]
acting in the so called extended Hilbert space which is, in our case,
\[
\sK=L^{2}((0,T)\times\mathbb{R}_{+},r\mbox{d}t\mbox{d}r).\]
The time derivative is taken with the periodic boundary conditions.
This approach, very similar to that usually applied in classical mechanics,
makes it possible to pass from a time-dependent system to an autonomous
one. The price to be paid for it is that one has to work with more
complex operators on the extended Hilbert space.

An important property of the quasienergy consists in its close relationship
to the Floquet operator \cite{Howland,Yajima}. In more detail, if
$\psi(t,r)\in\sK$ is an eigenfunction or a generalized eigenfunction
of $K$, $K\psi=\eta\psi$, which also implies that $\psi(t+T,r)=\psi(t,r)$,
then the wavefunction $e^{-i\eta t/\hbar}\psi(t,r)$ solves the Schr\"odinger
equation with the initial condition $\psi_{0}(r)=\psi(0,r)$. It follows
that $U(T,0)\psi_{0}=e^{-i\eta T/\hbar}\psi_{0}$. Thus from the spectral
decomposition of the quasienergy one can deduce the spectral decomposition
of the Floquet operator.

Let \[
K_{0}=-i\hbar\partial_{t}+H(p)\]
be the unperturbed quasienergy operator. Its complete set of normalized
eigenfunctions is \[
\{T^{-1/2}e^{im\Omega t}\varphi_{n}(p;r);\, m\in\mathbb{Z},n\in\mathbb{Z}_{+}\}\]
(here $\mathbb{Z}_{+}=\{0,1,2,\ldots\}$ stands for nonnegative integers,
the wave functions $\varphi_{n}(p;r)$ are defined in (\ref{eq:basis_varphi_n})),
with the corresponding eigenvalues $m\hbar\Omega+E_{n}(p)$. Thus
$K_{0}$ has a pure point spectrum which is in the resonant case (\ref{eq:resonance})
infinitely degenerated.

To take into account these degeneracies we perform the following transformation
of indices. Denote by $[x]$ and $\{x\}$ the integer and the fractional
part of a real number $x$, respectively, i.e. $x=[x]+\{x\}$, $[x]\in\mathbb{Z}$
and $0\leq\{x\}<1$. Furthermore, let \[
\rho(\mu,k)=\mu\,\{k/\mu\}\]
be the remainder in division of an integer $k$ by $\mu$. The transformation
of indices is a one-to-one map of $\mathbb{Z}\times\mathbb{Z}_{+}$
onto itself sending $(m,n)$ to $(k,\ell)$, with \begin{equation}
k=k(m,n):=\mu m+n,\ \ell=\ell(m,n):=[n/\mu],\label{eq:indices_kl_mn}\end{equation}
and, conversely,\begin{equation}
m=m(k,\ell):=[k/\mu]-\ell,\ n=n(k,\ell):=\mu\ell+\rho(\mu,k).\label{eq:indices_mn_kl}\end{equation}

Using the new indices $(k,\ell)$ we put\begin{equation}
\Psi_{k,\ell}(p;t,r)=T^{-1/2}\, e^{im(k,\ell)\Omega t}\,\varphi_{n(k,\ell)}(p;r).\label{eq:basis_Psi_kl}\end{equation}
Then the vectors $\Psi_{k,\ell}$, $(k,\ell)\in\mathbb{Z}\times\mathbb{Z}_{+}$,
form an orthonormal basis in the extended Hilbert space $\sK$. For
a fixed integer $k\in\mathbb{Z}$ let $P_{k}$ be the orthogonal projection
onto the subspace in $\sK$ spanned by the vectors $\Psi_{k,\ell}$,
$\ell\in\mathbb{Z}_{+}$. Then \begin{equation}
K_{0}=\sum_{k\in\mathbb{Z}}\lambda_{k}P_{k}\text{ where }\lambda_{k}=\hbar\omega_{c}(k+p+1/2).\label{eq:def_K0}\end{equation}

Furthermore, using the basis $\{\Psi_{k,\ell}\}$ one can identify
$\sK$ with the Hilbert space $\ell^{2}(\mathbb{Z}\times\mathbb{Z}_{+})$.
In particular, partial differential operators in the variables $t$
and $r$ like the quasienergy are identified in this way with matrix
operators. In the sequel we denote matrix operators by bold uppercase
letters.

\section{The quantum averaging method}

The full quasienergy operator $K=K(\epsilon)$ depends on the small
parameter $\epsilon$. Let us write $K(\epsilon)$ as a formal power
series, $K(\epsilon)=K_{0}+\epsilon K_{1}+\epsilon^{2}K_{2}+\ldots$.
In our case,\begin{equation}
K_{1}=f(\Omega t)\hbar\omega_{c}\!\left(\frac{\hbar p}{M\omega_{c}r^{2}}+\frac{1}{2}\right)\!,\ K_{2}=\frac{f(\Omega t)^{2}\hbar^{2}}{2Mr^{2}}\,,\label{eq:def_K1_K2}\end{equation}
and $K_{3}=K_{4}=\ldots=0$. The ultimate goal of the quantum averaging
method in the case of resonances is a unitary transformation resulting
in a partial (block-wise) diagonalization of $K(\epsilon)$. Thus
one seeks a skew-Hermitian operator $W(\epsilon)$ so that $e^{W(\epsilon)}K(\epsilon)e^{-W(\epsilon)}$
commutes with $K_{0}$ which is the same as saying that it commutes
with all projections $P_{k}$. This goal is achievable in principle
through an infinite recurrence which in practice should be interrupted
at some step. Here we shall be content with the first order approximation.

Let us introduce the block-wise diagonal part of an operator $A$
in $\sK$ as \[
\diag A:=\sum_{k\in\mathbb{Z}}P_{k}AP_{k}.\]
Thus $\diag A$ surely commutes with $K_{0}$. The off-diagonal part
is then defined as $\offdiag A:=A-\diag A$. Developing formally in
$\epsilon$ one has $W(\epsilon)=\epsilon W_{1}+O(\epsilon^{2})$
and\[
e^{W(\epsilon)}K(\epsilon)e^{-W(\epsilon)}=K_{0}+\epsilon K_{1}+\epsilon\,[W_{1},K_{0}]+O(\epsilon^{2}).\]
Choosing $W_{1}$ as \[
W_{1}=\sum_{k_{1},k_{2},k_{1}\neq k_{2}}(\lambda_{k_{1}}-\lambda_{k_{2}})^{-1}P_{k_{1}}K_{1}P_{k_{2}}\]
one has \begin{equation}
[W_{1},K_{0}]=-\offdiag K_{1}\label{eq:comm_W1_K0}\end{equation}
and \[
e^{W(\epsilon)}K(\epsilon)e^{-W(\epsilon)}=K_{0}+\epsilon\diag K_{1}+O(\epsilon^{2}).\]

Let us note that the solution is also expressible in terms of averaging
integrals, and this explains the name of the method \cite{Scherer_I,JauslinGuerinThomas}.
In more detail, one has\begin{equation}
\diag A=\lim_{\tau\to\infty}\,\frac{1}{\tau}\int_{0}^{\tau}e^{-iuK_{0}/\hbar}A\, e^{iuK_{0}/\hbar}\,\mbox{d}u\label{eq:diagA_int}\end{equation}
and\begin{equation}
W_{1}=\lim_{\tau\to\infty}\,\frac{i}{\hbar\tau}\int_{0}^{\tau}(\tau-u)e^{-iuK_{0}/\hbar}\offdiag(K_{1})e^{iuK_{0}/\hbar}\,\mbox{d}u.\label{eq:W1_int}\end{equation}

After switching on the perturbation, any unperturbed eigenvalue $\lambda_{k}$
gives rise to a perturbed spectrum which, in the first order approximation,
equals the spectrum of the operator $\lambda_{k}P_{k}+\epsilon P_{k}K_{1}P_{k}$
restricted to the subspace $\Ran P_{k}\subset\sK$. If the degeneracy
of $\lambda_{k}$ is infinite then the character of the perturbed
spectrum may be arbitrary, depending on the properties of $P_{k}K_{1}P_{k}$.
The corresponding perturbed (generalized) eigenvectors span a subspace
which is the range of the orthogonal projection\begin{eqnarray*}
P_{k}(\epsilon) & := & e^{-W(\epsilon)}P_{k}e^{W(\epsilon)}\,=\, P_{k}-\epsilon\,[W_{1},P_{k}]+O(\epsilon^{2})\\
 & = & P_{k}-\epsilon\,(\hat{S}_{k}K_{1}P_{k}+P_{k}K_{1}\hat{S}_{k})+O(\epsilon^{2})\end{eqnarray*}
where\[
\hat{S}_{k}=\sum_{\ell,\ell\neq k}(\lambda_{\ell}-\lambda_{k})^{-1}P_{\ell}\]
is the reduced resolvent of $K_{0}$ taken at the isolated eigenvalue
$\lambda_{k}$. Thus the first order averaging method is in fact nothing
but the standard quantum perturbation method in the first order but
accomplished on the extended Hilbert space simultaneously for all
eigenvalues of $K_{0}$ (compare to \cite[Chp.~II\S2]{Kato}).

Our strategy in the remainder of the paper is based on replacing the
true quasienergy $K(\epsilon)$ by its first order approximation \begin{equation}
K_{(1)}:=K_{0}+\epsilon\diag K_{1}\label{eq:def_K_trunc1}\end{equation}
and, consequently, $U(T,0)$ is replaced by an approximate Floquet
operator $U_{(1)}$ associated with $K_{(1)}$. To determine the approximate
Floquet operator $U_{(1)}$ one has to solve the spectral problem
for $K_{(1)}$. To this end, as already pointed out above, one can
employ the orthonormal basis $\{\Psi_{k\ell}\}$ in order to identify
operators in $\sK$ with infinite matrices indexed by $\mathbb{Z}\times\mathbb{Z}_{+}$.

Let $\{e_{k}^{1};k\in\mathbb{Z}\}$ denote the standard basis in $\ell^{2}(\mathbb{Z})$,
and $\{e_{\ell}^{2};\ell\in\mathbb{Z}_{+}\}$ denote the standard
basis in $\ell^{2}(\mathbb{Z}_{+})$. It is convenient to write $\ell^{2}(\mathbb{Z}\times\mathbb{Z}_{+})$
as the tensor product of Hilbert spaces $\ell^{2}(\mathbb{Z})\otimes\ell^{2}(\mathbb{Z}_{+})$
which also means identification of the standard basis in $\ell^{2}(\mathbb{Z}\times\mathbb{Z}_{+})$
with the set of vectors $\{e_{k}^{1}\otimes e_{\ell}^{2};\, k\in\mathbb{Z},\ell\in\mathbb{\mathbb{Z}}_{+}\}$.

Let $\boldsymbol{P}_{k}$ be the orthogonal projection onto the one-dimensional
subspace $\mathbb{C}e_{k}^{1}\subset\ell^{2}(\mathbb{Z})$. Recalling
(\ref{eq:def_K_trunc1}), (\ref{eq:def_K0}) and (\ref{eq:def_K1_K2}),
the matrix $\boldsymbol{K}_{(1)}$ of the operator $K_{(1)}$ expressed
in the basis (\ref{eq:basis_Psi_kl}) takes the form \begin{equation}
\boldsymbol{K}_{(1)}=\sum_{k\in\mathbb{Z}}\boldsymbol{P}_{k}\otimes(\lambda_{k}+\epsilon\boldsymbol{A}_{k})\label{eq:K1aver_matrix}\end{equation}
where $\boldsymbol{A}_{k}$ is the matrix operator in $\ell^{2}(\mathbb{Z}_{+})$
with the entries \begin{equation}
(\boldsymbol{A}_{k})_{\ell_{1},\ell_{2}}=\left\langle \,\Psi_{k,\ell_{1}},K_{1}\Psi_{k,\ell_{2}}\,\right\rangle _{\sK}.\label{eq:Ak_matrix}\end{equation}

To compute the matrix entries of $\boldsymbol{A}_{k}$ one observes
that formally (see (\ref{eq:def_H_p}))\begin{equation}
K_{1}=f(\Omega t)\,\partial H(p)/\partial p\label{eq:K1_eq_f_derH}\end{equation}
and so\[
\left\langle \,\Psi_{k,\ell_{1}},K_{1}\Psi_{k,\ell_{2}}\,\right\rangle _{\sK}=\sF[f](\ell_{2}-\ell_{1})\,\left\langle \varphi_{n(k,\ell_{1})}(p),\left(\partial H(p)/\partial p\right)\varphi_{n(k,\ell_{2})}(p)\right\rangle \]
where \[
\sF[f](j)=(2\pi)^{-1}\int_{0}^{2\pi}e^{-ij\cdot s}f(s)\,\mathrm{d}s\]
stands for the $j$th Fourier coefficient of $f$. Recall that, by
the assumption (\ref{eq:aver_f_0}), $\sF[f](0)=0$. Moreover, for
$\ell_{1}\neq\ell_{2}$ one has $n(k,\ell_{1})\neq n(k,\ell_{2})$,
hence \begin{equation}
\left\langle \varphi_{n(k,\ell_{1})}(p),H(p)\varphi_{n(k,\ell_{2})}(p)\right\rangle =0.\label{eq:scalar_varphi_H_varphi_0}\end{equation}
In \cite{AschHradeckyStovicek} it is derived that, for $n_{1}\neq n_{2}$,\begin{equation}
\left\langle \varphi_{n_{1}}(p),\frac{\partial\varphi_{n_{2}}(p)}{\partial p}\right\rangle =\frac{1}{2(n_{2}-n_{1})}\,\min\!\left\{ \frac{\gamma(p;n_{2})}{\gamma(p;n_{1})}\,,\frac{\gamma(p;n_{1})}{\gamma(p;n_{2})}\right\} \label{eq:scalar_varphi_der_psi}\end{equation}
where \[
\gamma(p;n)=\big(\Gamma(n+p+1)/n!\big)^{1/2}.\]
Differentiating (\ref{eq:scalar_varphi_H_varphi_0}) with respect
to $p$ and using (\ref{eq:scalar_varphi_der_psi}) one finally obtains
the relation\begin{equation}
(\boldsymbol{A}_{k})_{\ell_{1},\ell_{2}}=\frac{\hbar\omega_{c}}{2}\,\sF[f](\ell_{2}-\ell_{1})\,\min\!\left\{ \frac{\gamma(p;n(k,\ell_{2}))}{\gamma(p;n(k,\ell_{1}))}\,,\frac{\gamma(p;n(k,\ell_{1}))}{\gamma(p;n(k,\ell_{2}))}\right\} .\label{eq:Ak_l1l2}\end{equation}

Note that $n(k,\ell)$, as defined in (\ref{eq:indices_mn_kl}), is
$\mu$--periodic in the integer variable $k$, and so is the matrix
$\boldsymbol{A}_{k}$, i.e. $\boldsymbol{A}_{k+\mu}=\boldsymbol{A}_{k}$.
Moreover, since $\mu\omega_{c}=2\pi/T$ one also has $e^{-i\lambda_{k+\mu}T/\hbar}=e^{-i\lambda_{k}T/\hbar}$
(see (\ref{eq:def_K0})). For an integer $s$, $0\leq s<\mu$, let
$\sH_{s}$ be the closed subspace in the original Hilbert space $\sH=L^{2}(\mathbb{R}_{+},r\mbox{d}r)$
spanned by the vectors $\varphi_{s+j\mu}(r)$, $j=0,1,2,\ldots$.
Then $\sH$ decomposes into the orthogonal sum \[
\sH=\sH_{0}\oplus\sH_{1}\oplus\ldots\oplus\sH_{\mu-1},\]
and from the relationship between $K_{(1)}$ and $U_{(1)}$, as recalled
in Section~\ref{sec:Floquet_quasienergy}, it follows that every
subspace $\sH_{s}$ is invariant with respect to $U_{(1)}$.

In the example which we study in more detail in the following section
(for a sinusoidal function $f(t)$), the matrix operators $\boldsymbol{A}_{s}$
have purely absolutely continuous spectra. For the sake of simplicity
of the notation let us confine ourselves to this case. For a fixed
index $s$, $0\leq s<\mu$, suppose that all generalized eigenvectors
and eigenvalues of $\boldsymbol{A}_{s}$ are parametrized by a parameter
$\theta\in(a_{s},b_{s})$. Let us call them $\boldsymbol{x}_{s}(\theta)$
and $\eta_{s}(\theta)$, respectively, i.e. \[
\boldsymbol{A}_{s}\boldsymbol{x}_{s}(\theta)=\eta_{s}(\theta)\boldsymbol{x}_{s}(\theta),\]
and write \[
\boldsymbol{x}_{s}(\theta)=(\xi_{s;0}(\theta),\xi_{s;1}(\theta),\xi_{s;2}(\theta),\ldots).\]
The generalized eigenvectors $\boldsymbol{x}_{s}(\theta)$ are supposed
to be normalized to the $\delta$ function, i.e. \[
\langle\boldsymbol{x}_{s}(\theta_{1}),\boldsymbol{x}_{s}(\theta_{2})\rangle=\delta(\theta_{1}-\theta_{2}),\]
which in fact means that $\xi_{s;\ell}(\theta)$ as a function in
the variables $\ell\in\mathbb{Z}_{+}$ and $\theta\in(a_{s},b_{s})$
is a kernel of a unitary mapping between the Hilbert spaces $\ell^{2}(\mathbb{Z}_{+})$
and $L^{2}((a_{s},b_{s}),\mbox{d}\theta)$. Thus the spectral decomposition
of $\boldsymbol{A}_{s}$ reads:\[
\forall\boldsymbol{v}\in\ell^{2}(\mathbb{Z}_{+}),\ \boldsymbol{A}_{s}\boldsymbol{v}=\int_{a_{s}}^{b_{s}}\eta_{s}(\theta)\langle\boldsymbol{x}_{s}(\theta),\boldsymbol{v}\rangle\,\boldsymbol{x}_{s}(\theta)\,\mbox{d}\theta.\]

Put \begin{equation}
\Xi_{s}(\theta,r)=\sum_{j=0}^{\infty}\xi_{s;j}(\theta)\,\varphi_{s+j\mu}(p;r).\label{eq:Xi_th_r_def}\end{equation}
Then again,\[
\int_{0}^{\infty}\overline{\Xi_{s}(\theta_{1},r)}\,\Xi_{s}(\theta_{2},r)\, r\mbox{d}r=\delta(\theta_{1}-\theta_{2})\]
and, for all $\psi(r)\in\sH_{s}$,\begin{equation}
U_{(1)}\psi(r)=e^{-2\pi i(s+p+1/2)/\mu}\int_{a_{s}}^{b_{s}}e^{-i\epsilon\,\eta_{s}(\theta)T/\hbar}\,\langle\,\Xi_{s}(\theta),\psi\rangle\,\Xi_{s}(\theta,r)\,\mbox{d}\theta.\label{eq:U_approx_def}\end{equation}

To get a correct approximation in the first order of the propagator
one further has to take into account the transformation which is inverse
to that generated by $W($$\epsilon)\approx\epsilon\, W_{1}$. First
observe that $W_{1}$ is a multiplication operator on the Hilbert
space $\sK$ in the following sense. Let $S$ be the unitary operator
on $\sK$ acting as\[
S\psi(t,r)=e^{i\Omega t}\psi(t,r),\ \forall\psi\in\sK.\]
An operator $L$ on $\sK$ commutes with $S$ if and only if there
exists a one-parameter $T$-periodic family of operators $\mathcal{L}(t)$
on $L^{2}(\mathbb{R}_{+},r\mbox{d}r)$ such that $L\psi(t,r)=\mathcal{L}(t)\psi(t,r)$.
Notice that \[
S^{-1}K_{0}S=K_{0}+\hbar\Omega.\]
With this equality, it is obvious from (\ref{eq:diagA_int}) that
if $A$ commutes with $S$ then the same is true for $\diag A$. Furthermore,
as one can see from (\ref{eq:def_K1_K2}), $K_{1}$ commutes with
$S$, and from (\ref{eq:W1_int}) one infers that $W_{1}$ commutes
with $S$ as well. Hence there exists a one-parameter $T$-periodic
family of skew-Hermitian operators $\mathcal{W}_{1}(t)$ on $L^{2}(\mathbb{R}_{+},r\mbox{d}r)$
such that \[
W_{1}\psi(t,r)=\mathcal{W}_{1}(t)\psi(t,r),\ \forall\psi\in\sK.\]

Next notice that a transformation of the quasienergy operator of the
form $\tilde{K}=e^{\mathcal{W}(t)}Ke^{-\mathcal{W}(t)}$, where again
$\mathcal{W}(t)$ is a $T$-periodic family of skew-Hermitian operators
on $L^{2}(\mathbb{R}_{+},r\mbox{d}r)$, implies a transformation of
the associated propagators according to the rule\[
\tilde{U}(t_{1},t_{2})=e^{\mathcal{W}(t_{1})}U(t_{1},t_{2})e^{-\mathcal{W}(t_{2})}.\]
Hence the correct approximation of the Floquet operator reads\begin{equation}
U(T,0)\approx U_{\text{approx}}=e^{-\epsilon\mathcal{W}_{1}(0)}U_{(1)}e^{\epsilon\mathcal{W}_{1}(0)}.\label{eq:Uapprox}\end{equation}
Let us note, however, that one has, for $N\in\mathbb{N}$ and $\psi\in\sK$,\begin{equation}
\langle U_{\text{approx}}^{\ N}\psi,H(p)U_{\text{approx}}^{\ N}\psi\rangle=\langle U_{(1)}^{\ N}\psi_{1},\left(H(p)+\epsilon\,[\mathcal{W}_{1}(0),H(p)]\right)U_{(1)}^{\ N}\psi_{1}\rangle+O(\epsilon^{2})\label{eq:Emean_approx}\end{equation}
where $\psi_{1}=e^{\epsilon\mathcal{W}_{1}(0)}\psi$. If the commutator
$[\mathcal{W}_{1}(0),H(p)]$ happens to be bounded then it does not
contribute to the acceleration rate.

Finally let us indicate how to compute the operator-valued function
$\mathcal{W}_{1}(t)$. One has (here $\varphi_{n}=\varphi_{n}(p;r)$)\begin{equation}
\mathcal{W}_{1}(t)=\sum_{j=-\infty}^{\infty}\,\,\sum_{n_{1},n_{2}=0}^{\infty}e^{i\Omega jt}\, w(j,n_{1},n_{2})\,\langle\varphi_{n_{2}},\cdot\,\rangle\,\varphi_{n_{1}}\label{eq:W1_Fourier}\end{equation}
where\[
w(j,n_{1},n_{2})=\frac{1}{T}\int_{0}^{T}e^{-i\Omega jt}\,\langle\varphi_{n_{1}},\mathcal{W}_{1}(t)\varphi_{n_{2}}\rangle\,\mbox{d}t.\]
The commutator equation (\ref{eq:comm_W1_K0}) is equivalent to the
differential equation\begin{equation}
-i\hbar\,\mathcal{W}_{1}\,'(t)+[H(p),\mathcal{W}_{1}(t)]=\offdiag K_{1}.\label{eq:W1_ODE}\end{equation}
Substituting (\ref{eq:W1_Fourier}) into (\ref{eq:W1_ODE}) and using
(\ref{eq:K1_eq_f_derH}) jointly with (\ref{eq:diagA_int}) one finds
that\begin{equation}
w(j,n_{1},n_{2})=\frac{\sF[f](j)}{\hbar\omega_{c}(\mu j+n_{1}-n_{2})}\left\langle \!\varphi_{n_{1}},\frac{\partial H(p)}{\partial p}\,\varphi_{n_{2}}\!\right\rangle \ \,\text{if}\ \,\mu j+n_{1}-n_{2}\neq0\label{eq:w_jn1n2}\end{equation}
and $w(j,n_{1},n_{2})=0$ otherwise.

\section{A sinusoidally time-dependent AB flux}

In the remainder of the paper we discuss the example when $f(t)=\sin(t)$.
The goal of the current section is to provide more details on the
spectral decomposition of the averaged quasienergy $K_{(1)}$ derived
in (\ref{eq:def_K_trunc1}). Naturally, rather than directly with
the quasienergy we shall deal with its matrix, as given in (\ref{eq:K1aver_matrix})
and (\ref{eq:Ak_matrix}).

We still assume that $s\in\{0,1,\ldots,\mu-1\}$ is fixed. For this
choice of $f(t)$, an immediate evaluation of formula (\ref{eq:Ak_l1l2})
gives\[
(\boldsymbol{A}_{s})_{j_{1},j_{2}}=\frac{\hbar\omega_{c}}{4i}\,\delta_{|j_{2}-j_{1}|,1}\sign(j_{2}-j_{1})\left(\prod_{\nu=1}^{\mu}\,\frac{\mu j_{<}+s+\nu}{\mu j_{<}+s+p+\nu}\right)^{\!1/2}\!.\]
where $j_{<}=\min\{j_{1},j_{2}\}$. Thus one has \[
\boldsymbol{A}_{s}=(\hbar\omega_{c}/4)\,\boldsymbol{D}\boldsymbol{J}\boldsymbol{D}^{-1}\]
where $\boldsymbol{J}$ is the Jacobi (tridiagonal) matrix with zero
diagonal,\begin{equation}
\boldsymbol{J}=\left(\begin{array}{ccccc}
0 & \alpha_{0} & 0 & 0 & \ldots\\
\alpha_{0} & 0 & \alpha_{1} & 0 & \ldots\\
0 & \alpha_{1} & 0 & \alpha_{2} & \ldots\\
0 & 0 & \alpha_{2} & 0 & \ldots\\
\vdots & \vdots & \vdots & \vdots & \ddots\end{array}\right)\!,\label{eq:J}\end{equation}
and with the positive entries \[
\alpha_{j}=\left(\prod_{\nu=1}^{\mu}\,\frac{\mu j+s+\nu}{\mu j+s+p+\nu}\right)^{\!1/2},\]
and $\boldsymbol{D}$ is the unitary diagonal matrix with the diagonal
$(1,i,i^{2},i^{3},\ldots)$.

This is an elementary fact that the spectrum of $\boldsymbol{J}$
is simple since any eigenvector or generalized eigenvector is unambiguously
determined by its first entry. Moreover, one readily observes that
the matrices $\boldsymbol{J}$ and $-\boldsymbol{J}$ are unitarily
equivalent, and so the spectrum of $\boldsymbol{J}$ is symmetric
with respect to the origin.

In our case, \[
\alpha_{j}=1-p/(2j)+O(j^{-2})\ \text{ as}\ j\to\infty.\]
Hence $\boldsymbol{J}$ is rather close to the {}``free'' Jacobi
matrix $\boldsymbol{J}_{0}$ for which $\alpha_{0,j}=1$ for all $j$.
The spectral problem for $\boldsymbol{J}_{0}$ is readily solvable
explicitly (see below). It turns out that the spectral properties
of $\boldsymbol{J}$ are close to those of $\boldsymbol{J}_{0}$ as
well \cite{JanasMoszynski}, see also \cite{Teschl}. In particular,
it is known that the singular continuous spectrum of $\boldsymbol{J}$
is empty, the essential spectrum coincides with the absolutely continuous
spectrum and equals the interval $[-2,2\,]$. Furthermore, there are
no embedded eigenvalues, i.e.\ if $\eta$ is an eigenvalue of $\boldsymbol{J}$
then $|\eta|\geq2$.

Splitting $\boldsymbol{J}$ into the sum of the upper triangular and
the lower triangular part, one notes that $\|\boldsymbol{J}\|\leq2\sup\{\alpha_{0},\alpha_{1},\alpha_{2},\ldots\}$.
In our example, $\alpha_{j}\leq1$ for all $j$ and so $\|\boldsymbol{J}\|\leq2$
and, consequently, the spectrum of $\boldsymbol{J}$ is contained
in the interval $[-2,2\,]$. This means that the only possible eigenvalues
of $\boldsymbol{J}$ are $\pm2$. But one can exclude even this possibility.
In fact, suppose that $\boldsymbol{J}\boldsymbol{u}=2\boldsymbol{u}$,
with $\boldsymbol{u}=(u_{0},u_{1},u_{2},\ldots)$ and $u_{0}=1$.
Then \[
\alpha_{j-1}u_{j-1}+\alpha_{j}u_{j+1}=2u_{j}\text{ for }j=0,1,2,\ldots\]
(while putting $u_{-1}=0$). Summing this equality for $j=0,1,\ldots,n$,
and using that $\alpha_{j}\leq1$, one finds that $u_{n+1}\geq u_{n}+1$
for $n=0,1,2,\ldots$. Hence $u_{j}\geq j+1$ for all $j$, and so
$\boldsymbol{u}$ is not square summable. Thus one can summarize that
the spectrum of $\boldsymbol{J}$ is simple, purely absolutely continuous
and equals $[-2,2\,]$.

Let us parametrize the spectrum of $\boldsymbol{A}_{s}=\boldsymbol{A}_{s}(p)$
by a continuous parameter $\theta$, $0<\theta<\pi$, so that \[
\eta(\theta):=(\hbar\omega_{c}/2)\cos(\theta)\]
is a point from the spectrum and $\boldsymbol{x}(p;\theta)$ is the
corresponding normalized generalized eigenvector with components $\xi_{j}(p;\theta)$,
$j=0,1,2,\ldots$ (here we drop the index $s$ at $\boldsymbol{x}$
and $\mathbb{\xi}$ in order to simplify the notation). The asymptotic
behavior of the components $\xi_{j}$ is known \cite{JanasNaboko,BelovRybkin};
one has\begin{equation}
\xi_{j}(p;\theta)\sim A(p;\theta)\, i^{j}\cos\!\big(j\theta-(p/2)\cot(\theta)\log(j+1)+\phi(p;\theta)\big)\label{eq:xi_j_asympt}\end{equation}
for $j\gg0$. Here $A(p;\theta)$ is a normalization constant and
$\phi(p;\theta)$ is a phase which depends on the initial conditions
imposed on the sequence $\{\xi_{j}\}$ (the initial condition is simply
$\xi_{-1}=0$) but the asymptotic methods employed in the cited articles
do not provide an explicit value for it. In the limit case $p=0$
the generalized eigenvectors are known explicitly, namely \[
\xi_{j}(0;\theta)=\sqrt{2/\pi}\, i^{j}\sin((j+1)\theta)\]
for all $j$. Hence $\phi(0;\theta)=\theta-\pi/2$.

The generalized eigenvectors are supposed to be normalized so that
\[
\langle\boldsymbol{x}(p;\theta_{1}),\boldsymbol{x}(p;\theta_{2})\rangle=\delta(\theta_{1}-\theta_{2}).\]
For $p=0$, one can use the equality \[
\sum_{n=1}^{\infty}e^{inx}=\pi\delta(x)-\mathcal{P}\frac{1}{1-e^{-ix}}\]
which is valid for $x=\theta_{1}-\theta_{2}\in(-\pi,\pi)$ and where
the symbol $\mathcal{P}$ indicates the regularization of a nonintegrable
singularity in the sense of the principal value. The normalization
is an immediate consequence of this identity.

For general $p$, the contribution to the $\delta$ function should
come from the most singular and, at the same time, the leading term
in the asymptotic expansion of $\xi_{j}(p;\theta)$, as given in (\ref{eq:xi_j_asympt}).
This time, when investigating the singularity near the diagonal $\theta_{1}=\theta_{2}$
in the scalar product of two generalized eigenvectors, one is lead
to considering the sum \[
\sum_{n=1}^{\infty}n^{iax}e^{inx}\]
where $a=p/(2\sin^{2}\theta_{1})$ is a real constant. Using the Lerch
function $\Phi(z,s,v)$ one has for $|z|<1$ (see \cite[\S~9.55]{GradshteynRyzhik}),\[
\sum_{n=1}^{\infty}n^{s}z^{n}=z\,\Phi(z,s,1)=\Gamma(1-s)\,\sum_{n=-\infty}^{\infty}\left(-\log(z)+2\pi ni\right)^{-1+s}\!.\]
From here one deduces that, for any real $a$,\begin{equation}
\sum_{n=1}^{\infty}n^{iax}e^{inx}=\pi\delta(x)+i\mathcal{P}\frac{1}{x}+g_{a}(x)\label{eq:sum_eq_pi_delta}\end{equation}
where $g_{a}(x)$ is a regular distribution, i.e.\ a locally integrable
function. Hence in the general case, too, the normalization constant
is given by \[
A(p;\theta)=\sqrt{2/\pi}\,.\]

As already mentioned, the phase $\phi(p;\theta)$ in the asymptotic
solution (\ref{eq:xi_j_asympt}) remains undetermined. But we remark
that a bit more can be said about the behavior of the phase near the
spectral point $0$ (the center of the spectrum) which corresponds
to the value of the parameter $\theta=\pi/2$. More precisely, one
can compute the derivative $\partial\phi(p;\pi/2)/\partial\theta$.
Though this result is not directly used in the sequel it represents
an additional information about generalized eigenfunctions of $\boldsymbol{J}$.
We briefly indicate basic steps of the computation in Appendix.

\section{The acceleration rate}

In the case when $f(t)=\sin(t)$ the commutator $[\mathcal{W}_{1}(0),H(p)]$
occurring in (\ref{eq:Emean_approx}) can be shown to be bounded.
This implies that instead of the approximate Floquet operator $U_{\text{approx}}$,
as given in (\ref{eq:Uapprox}), one can work directly with $U_{(1)}$
defined in (\ref{eq:U_approx_def}) when deriving a formula for the
acceleration rate. On the other hand, one should not forget about
the transformation of the initial state, i.e. $\psi_{0}$ has to be
replaced by $e^{\epsilon\mathcal{W}_{1}(0)}\psi_{0}$, see (\ref{eq:Emean_approx}).

First let us shortly discuss the boundedness of the commutator. From
(\ref{eq:W1_Fourier}) and (\ref{eq:w_jn1n2}) while using also (\ref{eq:scalar_varphi_der_psi})
one derives that for $n_{1},n_{2}\in\mathbb{Z}_{+}$, $n_{1}-n_{2}\neq\pm\mu$,\begin{align}
 & \langle\varphi_{n_{1}}(p),[H(p),\mathcal{W}_{1}(0)]\varphi_{n_{2}}(p)\rangle\nonumber \\
\noalign{\medskip} & \quad=\,\frac{\hbar\omega_{c}}{4i}\,\frac{n_{1}-n_{2}}{(n_{1}-n_{2})^{2}-\mu^{2}}\,\min\!\left\{ \frac{\gamma(p;n_{2})}{\gamma(p;n_{1})}\,,\frac{\gamma(p;n_{1})}{\gamma(p;n_{2})}\right\} \!.\label{eq:commut_matrix}\end{align}
Of course, the parallels to the diagonal determined by $n_{1}-n_{2}=\pm\mu$
can be explicitly evaluated as well but for our purposes it is sufficient
to know that they are bounded. In \cite[Lemma~6]{AschHradeckyStovicek}
it is shown that the matrix operator \textbf{$\boldsymbol{Q}$} in
$\ell^{2}(\mathbb{Z}_{+})$ with the entries\begin{equation}
\boldsymbol{Q}_{n_{1},n_{2}}=\frac{\hbar\omega_{c}}{4i\,(n_{1}-n_{2})}\,\min\!\left\{ \frac{\gamma(p;n_{2})}{\gamma(p;n_{1})}\,,\frac{\gamma(p;n_{1})}{\gamma(p;n_{2})}\right\} \label{eq:Q_matrix}\end{equation}
for $n_{1}\neq n_{2}$ and $0$ otherwise is bounded. Thus to verify
the boundedness of the commutator it suffices to show that the difference
of matrices (\ref{eq:commut_matrix}) and (\ref{eq:Q_matrix}) has
a finite operator norm. This can be readily done, for example, with
the aid of the following estimate for the norm of a Hermitian matrix
operator $\boldsymbol{B}$ \cite[\S~I.4.3]{Kato},\[
\|\boldsymbol{B}\|\leq\sup_{n_{1}\in\mathbb{Z}_{+}}\,\sum_{n_{2}=0}^{\infty}|\boldsymbol{B}_{n_{1},n_{2}}|.\]

To proceed further, we again fix an integer $s$, $0\leq s<\mu$.
Suppose one is given a function $\varrho(\theta)\in C_{0}^{\infty}((0,\pi))$.
Recalling (\ref{eq:Xi_th_r_def}) we put \begin{equation}
\psi(r)=\int_{0}^{\pi}\Xi_{s}(\theta,r)\varrho(\theta)\,\mbox{d}\theta.\label{eq:varphi_int_varrho}\end{equation}
In what follows, we drop the index $s$ and, whenever convenient,
write simply $H$ instead of $H(p)$. Using (\ref{eq:U_approx_def}),
one has, for $N\in\mathbb{N}$,\begin{align*}
\langle U_{(1)}^{\,\, N}\psi,HU_{(1)}^{\,\, N}\psi\rangle & =\,\int_{0}^{\pi}\!\int_{0}^{\pi}e^{i\epsilon(\cos\theta_{1}-\cos\theta_{2})\omega_{c}TN/2}\\
\noalign{\smallskip} & \qquad\qquad\times\!\langle\,\Xi(\theta_{1}),H\Xi(\theta_{2})\rangle\,\overline{\varrho(\theta_{1})}\varrho(\theta_{2})\,\mbox{d}\theta_{1}\mbox{d}\theta_{2}\\
\noalign{\medskip} & =\,\sum_{j=0}^{\infty}E_{_{s+j\mu}}(p)\left|\int_{0}^{\pi}e^{-i\epsilon\cos(\theta)\omega_{c}TN/2}\xi_{j}(p;\theta)\varrho(\theta)\,\mbox{d}\theta\right|^{2}\!.\end{align*}
Note that $\{\xi_{j}(p;\theta)\}_{j=0}^{\infty}$ is an orthonormal
basis in $L^{2}((0,\pi),\mbox{d}\theta)$ and so \[
\sum_{j=0}^{\infty}\left|\int_{0}^{\pi}e^{-i\epsilon\cos(\theta)\omega_{c}TN/2}\xi_{j}(p;\theta)\varrho(\theta)\,\mbox{d}\theta\right|^{2}=\int_{0}^{\pi}|\varrho(\theta)|^{2}\,\mbox{d}\theta.\]
Hence, in view of (\ref{eq:E_n_p}), the leading contribution to the
acceleration rate comes from the expression\[
\mu\hbar\omega_{c}\sum_{j=0}^{\infty}(j+1)\left|\int_{0}^{\pi}e^{-i\epsilon\cos(\theta)\omega_{c}TN/2}\xi_{j}(p;\theta)\varrho(\theta)\,\mbox{d}\theta\right|^{2}.\]
Furthermore, restricting this sum to an arbitrarily large but finite
number of summands results in an expression which is uniformly bounded
in $N$. This justifies replacement of $\xi_{j}(p;\theta)$ by the
leading asymptotic term, as given in (\ref{eq:xi_j_asympt}) (with
$A(p;\theta)=\sqrt{2/\pi}$). Hence the leading contribution to the
acceleration rate is expressible as\[
\frac{2\hbar\Omega}{\pi}\,\int_{0}^{\pi}\!\int_{0}^{\pi}h(\theta_{1},\theta_{2})e^{i\epsilon(\cos\theta_{1}-\cos\theta_{2})\omega_{c}TN/2}\,\overline{\varrho(\theta_{1})}\varrho(\theta_{2})\,\mbox{d}\theta_{1}\mbox{d}\theta_{2}\]
where\begin{align}
h(\theta_{1},\theta_{2}) & =\sum_{j=0}^{\infty}\,(j+1)\cos\!\left(j\theta_{1}-\frac{p}{2}\,\cot(\theta_{1})\log(j+1)+\phi(p;\theta_{1})\right)\nonumber \\
 & \qquad\quad\times\,\cos\!\left(j\theta_{2}-\frac{p}{2}\,\cot(\theta_{2})\log(j+1)+\phi(p;\theta_{2})\right).\label{eq:h_th1_th2}\end{align}

The singular part of the distribution $h(\theta_{1},\theta_{2})$
is supported on the diagonal $\theta_{1}=\theta_{2}$. The sum in
(\ref{eq:h_th1_th2}) can be evaluated analogously as that in (\ref{eq:sum_eq_pi_delta})
with the result\begin{align*}
h(\theta_{1},\theta_{2})=\, & -\frac{1}{2}\,\frac{\partial}{\partial\theta_{2}}\mathcal{P}\frac{1}{\theta_{1}-\theta_{2}}-\frac{\pi}{2}\!\left(\frac{\partial\phi(p;\theta_{1})}{\partial\theta}-1\right)\!\delta(\theta_{1}-\theta_{2})\\
\noalign{\smallskip} & +\,\text{a regular distribution}.\end{align*}
Estimating the acceleration rate we can restrict ourselves to a sufficiently
small but fixed neighborhood of the diagonal with a radius $d>0$.
Thus we arrive at the expression \[
\frac{\hbar\Omega}{\pi}\,\mathcal{P}\!\underset{\substack{\mbox{\ensuremath{\phantom{{s}}}}\\
|\theta_{1}-\theta_{2}|<d}
}{\int_{0}^{\pi}\!\int_{0}^{\pi}}\frac{1}{\theta_{1}-\theta_{2}}\,\frac{\partial}{\partial\theta_{2}}\!\left(e^{-i\epsilon\sin(\theta_{1})(\theta_{1}-\theta_{2})\omega_{c}TN/2}\,\,\overline{\varrho(\theta_{1})}\varrho(\theta_{2})\right)\!\mbox{d}\theta_{1}\mbox{d}\theta_{2}.\]
Further we carry out the differentiation, as indicated in the integrand,
and get rid of the terms which are not proportional to $N$ or which
are non-singular. Moreover, we use the substitution $\theta_{2}=\theta_{1}+u$.
Thus we obtain the expression\begin{align*}
 & -\frac{i\epsilon\hbar\Omega\omega_{c}TN}{2\pi}\int_{0}^{\pi}\sin(\theta_{1})|\varrho(\theta_{1})|^{2}\left(\mathcal{P}\int_{-d}^{d}\frac{1}{u}\, e^{i\epsilon\sin(\theta_{1})\omega_{c}TNu/2}\,\mbox{d}u\right)\!\mbox{d}\theta_{1}\\
 & =\frac{\epsilon\hbar\Omega\omega_{c}TN}{\pi}\int_{0}^{\pi}\sin(\theta)|\varrho(\theta)|^{2}\!\left(\,\int_{0}^{d}\frac{1}{u}\sin\!\Big(\,\frac{\epsilon}{2}\sin(\theta)\omega_{c}TNu\Big)\mbox{d}u\!\right)\!\mbox{d}\theta.\end{align*}
Finally note that, for any $a$ real, \[
\lim_{N\to\infty}\int_{0}^{d}\frac{1}{u}\,\sin\!\left(aNu\right)\mbox{d}u=\frac{\pi}{2}\,\sign a.\]

Suppose that the initial state is chosen as $e^{-\epsilon\mathcal{W}_{1}(0)}\psi$.
Then we conclude that the formula for the acceleration rate in the
first-order  approximation reads\begin{eqnarray}
\gamma_{\text{acc}} & := & \lim_{N\to\infty}\left\langle U_{(1)}^{\, N}\psi,H(p)U_{(1)}^{\, N}\psi\right\rangle \Big/(NT\|\psi\|^{2})\nonumber \\
\noalign{\medskip} & = & \frac{|\epsilon|\hbar\omega_{c}\Omega}{2}\int_{0}^{\pi}\sin(\theta)|\varrho(\theta)|^{2}\,\mbox{d}\theta\bigg/\!\int_{0}^{\pi}|\varrho(\theta)|^{2}\,\mbox{d}\theta.\label{eq:acc_rate_quant}\end{eqnarray}
Here we have used that \[
\|\psi\|^{2}=\int_{0}^{\pi}|\varrho(\theta)|^{2}\,\mbox{d}\theta.\]

Formula (\ref{eq:acc_rate_quant}) can be compared to formula (\ref{eq:acc_rate_class}),
as derived for a classical particle, in the case when $\Phi(t)$ is
given by (\ref{eq:Phi_t}) and $f(t)=\sin(t)$. Then (\ref{eq:acc_rate_class})
gives the acceleration rate \[
\gamma_{\text{acc}}=|\epsilon|\hbar\omega_{c}\Omega\sin(\xi)/2\]
where $\xi\in(0,\pi)$ depends on some data which can be learned from
the asymptotic behavior of the classical trajectory. Let us finally
note that, according to the analysis and discussion of the classical
case presented in \cite{AschKalvodaStovicek}, the first-order averaging
approximation may in fact yield the correct acceleration rate (valid
for the original system), and this is so even if the parameter $\epsilon$
is not necessarily assumed to be very small.

\section{A numerical test}

We conclude our discussion by a presentation of a numerical result
that agree quite nicely with the predicted acceleration rate (\ref{eq:acc_rate_quant}).
For the sake of simplicity we put $\Omega=\omega_{c}=1$, and so $\mu=1$
and $s=0$. We still assume that $f(t)=\sin(t)$. Concerning the physical
constants, we set $\hbar=1$, $e=1$ and $M=1$. Furthermore, we choose
$p=2.5$, $\epsilon=0.4$, and for the density $\varrho(\theta)$
determining an initial state according to (\ref{eq:varphi_int_varrho})
we take the Gaussian function\[
\varrho(\theta)=\left(\frac{20}{\pi}\right)^{\!1/4}\exp\!\big(-10(2-\theta)^{2}+8i\theta\,\big)\]
restricted to the interval $\theta\in(0,\pi)$. Its values near the
limit points of the interval are in fact numerically indistinguishable
from $0$. Particularly, $\varrho(\theta)$ is normalized to unity
with a negligible error, i.e.\[
\int_{0}^{\pi}|\varrho(\theta)|^{2}\,\mbox{d}\theta=1.\]

The numerical method we use is based on expanding a solution of the
time-dependent Schr\"odinger equation with respect to the time-dependent
basis $\{\varphi_{n}(a(t));\, n\in\mathbb{Z}_{+}\}$, with $\varphi_{n}(p)$
being defined in (\ref{eq:basis_varphi_n}). Below we call the solution
of the Schr\"odinger equation $\psi(t)$. Recalling (\ref{eq:Xi_th_r_def})
we put \[
\psi_{0}(r)=\int_{0}^{\pi}\Xi_{0}(\theta,r)\varrho(\theta)\,\mathrm{d}\theta,\]
and we have $\|\psi_{0}\|=1$. The task is to solve the Cauchy problem
for the time-dependent Schr\"odinger equation \[
i\partial_{t}\psi(t)=H(a(t))\psi(t),\ \psi(0)=\tilde{\psi}_{0}:=e^{-\epsilon\mathcal{W}_{1}(0)}\psi_{0}.\]
Let us note that in the case of $f(t)=\sin(t)$ the matrix entries
of $\mathcal{W}_{1}(0)$ are expressed as the finite sum\[
\langle\varphi_{n_{1}}(p),\mathcal{W}_{1}(0)\varphi_{n_{2}}(p)\rangle=w(1,n_{1},n_{2})+w(-1,n_{1},n_{2}),\]
with $w(j,n_{1},n_{2})$ being given in (\ref{eq:w_jn1n2}) (with
$\mu=1$). To carry out the computations we truncate the Fourier expansion
of $\psi(t)$, \[
\psi(t)=\sum_{n=0}^{\infty}x_{n}(t)\varphi_{n}(a(t)),\ x_{n}(t)=\langle\varphi_{n}(a(t)),\psi(t)\rangle,\ n=0,1,\ldots,\]
at some fixed order $n_{\mathrm{max}}$. In this way we obtain a system
of ordinary differential equations for the Fourier coefficients \begin{align*}
ix'_{n}(t) & =E_{n}(a(t))x_{n}(t)-ia'(t)\sum_{j=0}^{n_{\mathrm{max}}}\,\langle\varphi_{n}(a(t)),\varphi'_{j}(a(t))\rangle\, x_{j}(t),\\
x_{n}(0) & =\langle\varphi_{n}(a(0)),\tilde{\psi}_{0}\rangle,\ n=0,1,\ldots,n_{\mathrm{max}}.\end{align*}
Explicit formulas for the scalar products are known from \cite{AschHradeckyStovicek}
(see (\ref{eq:scalar_varphi_der_psi})). In order to approximately
solve this system we employ the explicit Runge-Kutta method of order
4 (RK4) with an adaptive step-size control, and we choose $n_{\mathrm{max}}=120$. 

From the computational point of view it is convenient to introduce
the mean value of energy at time $t$ as \[
\mathcal{E}(t):=\langle\psi(t),H(a(t))\psi(t)\rangle.\]
$\mathcal{E}(t)$ is then approximated by the sum\[
\mathcal{E}(t)\approx\sum_{n=0}^{n_{\mathrm{max}}}E_{n}(a(t))|x_{n}(t)|^{2}.\]
The acceleration rate is computed according to formula (\ref{eq:acc_rate_quant})
in which one has to substitute $\psi_{0}$ for $\psi$. Let us point
out that this formula depends only on the time evolution over the
intervals which are integer multiples of the period $T$, and clearly,
$H(a(NT))=H(p)$ for $N=0,1,2,\ldots$. The predicted acceleration
rate for the above particular values of parameters is $\gamma_{\text{acc}}=0.1796$.
The numerically computed function $\mathcal{E}(t)/t$ is compared
to this value in Fig.~1.

\setcounter{equation}{0}
\renewcommand{\theequation}{A.\arabic{equation}}

\section*{Appendix. The phase $\phi(p;\theta)$ near the spectral point $0$}

Here we compute the derivative $\partial\phi(p;\pi/2)/\partial\theta$
of the phase $\phi(p;\theta)$ introduced in (\ref{eq:xi_j_asympt}).
We know that $0$ always belongs to the spectrum of the Jacobi matrix
$\boldsymbol{J}$ introduced in (\ref{eq:J}). Putting $\boldsymbol{u}=(u_{0},u_{1},u_{2},\ldots)$,
with $u_{2j+1}=0$ and\begin{equation}
u_{2j}=(-1)^{j}\,\prod_{k=0}^{j-1}\frac{\alpha_{2k}}{\alpha_{2k+1}}\label{eq:u_2j}\end{equation}
for $j=0,1,2,\ldots$, one has $\boldsymbol{J}\boldsymbol{u}=\boldsymbol{0}$
and $u_{0}=1$. Recalling that, in our example, $\alpha_{j}=1-p/(2j)+O(j^{-2})$
one derives that \[
u_{2j}=(-1)^{j}u_{\infty}\big(1+p/(8j)+O(j^{-2})\big)\quad\text{as\ }j\to\infty,\]
where \[
u_{\infty}=\lim_{j\to\infty}(-1)^{j}u_{2j}\]
is a finite constant (depending on $p$, however). Comparing to (\ref{eq:xi_j_asympt}),
with $A(p;\theta)=\sqrt{2/\pi}$ and $\theta=\pi/2$, one finds that
\[
\boldsymbol{x}(p;\pi/2)=(\sqrt{2/\pi}/u_{\infty})\,\boldsymbol{u}.\]
Moreover, $\phi(p;\pi/2)=0$.

Differentiating the equality \[
\boldsymbol{J}\boldsymbol{x}(p;\theta)=2\cos(\theta)\boldsymbol{x}(p;\theta)\]
with respect to $\theta$ at the point $\pi/2$ and using the substitution\[
\partial\boldsymbol{x}(p;\pi/2)/\partial\theta=-\left(2\sqrt{2/\pi}\big/u_{\infty}\right)\!\boldsymbol{v},\]
with $\boldsymbol{v}=(v_{0},v_{1},v_{2},\ldots)$, one arrives at
the equation $\boldsymbol{J}\boldsymbol{v}=\boldsymbol{u}$. From
(\ref{eq:xi_j_asympt}) one deduces that\begin{equation}
v_{j}\sim\frac{1}{2}\, u_{\infty}\sin\!\left(j\,\frac{\pi}{2}\right)\!\left(j+\frac{p}{2}\,\log(j+1)+\frac{\partial\phi(p;\pi/2)}{\partial\theta}\right)\label{eq:v_j_asympt}\end{equation}
for $j\gg0$. This suggests that one can seek a solution $\boldsymbol{v}$
such that $v_{2j}=0$ for all $j$. This assumption on $\boldsymbol{v}$
is in fact necessary and makes the solution unambiguous since otherwise
one could add to $\boldsymbol{v}$ any nonzero multiple of $\boldsymbol{u}$
thus violating the asymptotic behavior (\ref{eq:v_j_asympt}). Given
that all odd elements of the vector $\boldsymbol{u}$ and all even
elements of $\boldsymbol{v}$ vanish the equation $\boldsymbol{J}\boldsymbol{v}=\boldsymbol{u}$
effectively reduces to a linear system with a lower triangular matrix
which is explicitly solvable. Using (\ref{eq:u_2j}) one can express
the solution as\begin{equation}
v_{2j+1}=\frac{1}{\alpha_{2j}u_{2j}}\,\sum_{k=0}^{j}(u_{2k})^{2},\ \ j=0,1,2,\ldots.\label{eq:v_2j_plus_1}\end{equation}

Noting that\[
\sum_{k=0}^{j}\left(1+\frac{p}{4k+2}\right)=j+1+\frac{p}{4}\big(\log(4j+4)+\gamma_{E}\big)+O(j^{-1}),\]
where $\gamma_{E}$ is the Euler constant, and that \[
(u_{2k})^{2}=u_{\infty}^{\,2}(1+p/(4k)+O(k^{-2}))\]
one derives\begin{align*}
\sum_{k=0}^{j}\left(\frac{u_{2k}}{u_{\infty}}\right)^{\!2}=\,\, & j+1+\frac{p}{4}\,\big(\log(4j+4)+\gamma_{E}\big)\\
 & +\,\sum_{k=0}^{\infty}\left(\!\left(\frac{u_{2k}}{u_{\infty}}\right)^{\!2}-1-\frac{p}{4k+2}\right)+O(j^{-1}).\end{align*}
Using (\ref{eq:v_2j_plus_1}) and comparing to (\ref{eq:v_j_asympt})
one finally arrives at the relation\[
\frac{\partial\phi(p;\pi/2)}{\partial\theta}=1+\frac{p}{2}\big(\log(2)+\gamma_{E}\big)+2\sum_{j=0}^{\infty}\!\left(\!\left(\frac{u_{2j}}{u_{\infty}}\right)^{\!2}-1-\frac{p}{4j+2}\right)\!.\]

\section*{Acknowledgments}

The authors wish to acknowledge gratefully partial support from the
following grants: Grant No.\ 201/09/0811 of the Czech Science Foundation
(P.\v{S}.), Grant No.\ LC06002 of the Ministry of Education of the
Czech Republic and Grant No.\ 202/08/H072 of the Czech Science Foundation
(T.K.).

\clearpage{}\includegraphics[bb=0bp 0bp 447bp 286bp,scale=0.8]{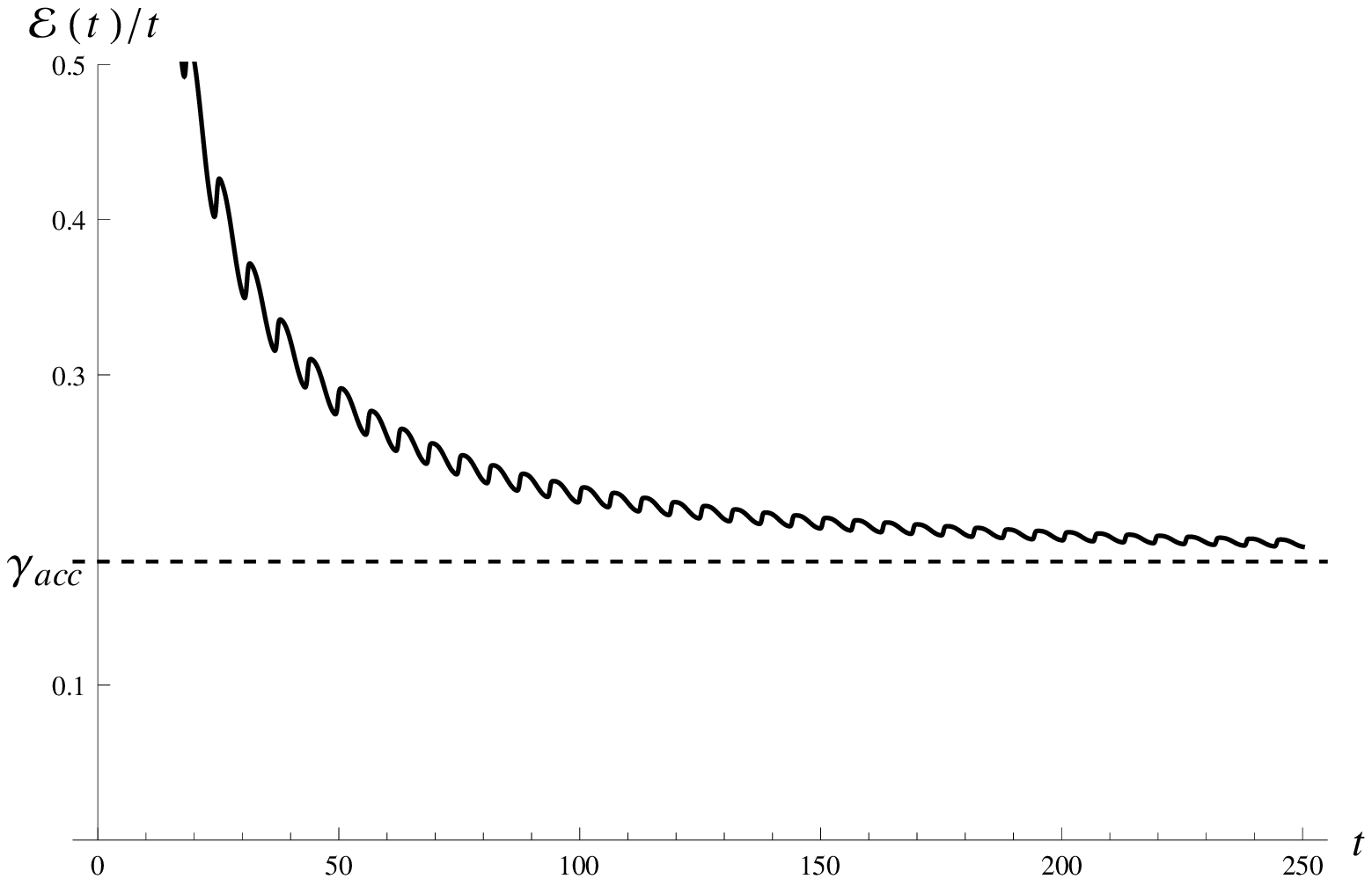}

\vspace{3\baselineskip}

\noindent Figure~1. The function $\mathcal{E}(t)/t$, with $\mathcal{E}(t)=\langle\psi(t),H(a(t))\psi(t)\rangle$
and $\psi(t)$ being a normalized solution of the time-dependent Schr\"odinger
equation, compared to the value of the acceleration rate $\gamma_{\text{acc}}$
derived in eq. (\ref{eq:acc_rate_quant}).
\end{document}